\input{aipcheck.tex}

\edef\optionlist{%
   \variorefoptionifavailable        
   draft,%
   \psnfssproblemoption              
   tnotealph}

\documentclass[\optionlist]{aipproc}

\layoutstyle{6x9}
\received{x}
\accepted{x}

\usepackage{ttct0001}

\usepackage{shortvrb}
\MakeShortVerb\|

\makeatletter
   \def\@oddfoot{\reset@font
                 \copyright{} 2004 AIP
                 \hfil\@title
                 \hfil\@date\hfil\thepage}
\makeatother

\begin{document}

\author{B. Bahr}{
  address={DAMTP, University of Cambridge,\\
  Wilberforce Road, Cambridge CB3 0WA, UK},
  email={bbahr@aei.mpg.de},
}

\author{B. Dittrich}{
  address={MPI f. Gravitational Physics, Albert Einstein Institute,\\
 \small Am M\"uhlenberg 1, D-14476 Potsdam, Germany},
  email={bdittrich@aei.mpg.de},
}

\title{Breaking and restoring of diffeomorphism symmetry in discrete gravity}
\date{}


\begin{abstract}
  We discuss the fate of diffeomorphism symmetry in discrete gravity. Diffeomorphism symmetry is typically broken by the discretization. This has repercussions for the observable content and the canonical formulation of the theory. It might however be possible to construct discrete actions, so--called perfect actions, with exact symmetries and we will review first steps towards this end.
\end{abstract}

\maketitle



\subsection{Introduction}


Systems with symmetry (either global or local gauge) very often lose that symmetry upon
discretization. While e.g. the introduction of a lattice in
Yang-Mills theory can be done without harming the local gauge
symmetry, it usually breaks invariance under global
Poincar\'{e}-transformation. In the same way the introduction of a
triangulation in Regge calculus \cite{regge} or in Spin Foam approaches \cite{spinfoams} breaks local diffeomorphism invariance. This is a problem for quantization as one might hope that keeping diffeomorphism invariance as far as possible might on the one hand avoid quantization ambiguities \cite{lost, ashtekarpaper} and on the other hand help to obtain general relativity (which is a  theory with diffeomorphism invariance) in the semi--classical limit \cite{kuchar}.

We will discuss whether discretization leads always to a breaking of diffeomorphism invariance or whether this can be avoided. The latter case would provide us a diffeomorphism invariant cut off method and hence be extremely valuable for the construction of quantum gravity theories.

\subsection{Reparametrization--invariant systems}

In this section we will discuss systems invariant under time reparametrization, which are essentially one--dimensional diffeomorphisms. Hence these systems share  certain features, that are essential to our discussion, with general relativity. Details about this discussion can be found in \cite{marsden, bahrdittrich2}.

Consider a system with one-dimensional configuration space, i.e.
with a coordinate $q$.  Assume a regular\footnote{ i.e. if $t'$ and $t''$ are
sufficiently close to each other, then for $'q$ and $q''$ there is a
unique solution $q(t)$ of the Euler-Lagrange equations with
$q(t')=q'$ and $q(t'')=q''$}
Lagrangian $L$, leading to the action
\begin{eqnarray}\label{Gl:OneDimensionalAction}
S\;=\;\int dt\;L(q(t),\dot q(t)),
\end{eqnarray}
\noindent where dot means derivative w.r.t. time $t$.

The dynamical system can be artificially enlarged by adding the time $t$ as another variable and
considering the evolution of $q,t$ w.r.t an auxiliary parameter $s$.
The \emph{extended} Lagrangian $L_e$ is given by
$L_e\big(q(s),t(s),q'(s),t'(s)\big):=L\big(q(s), q'(s)/t'(s)\big)\;t'(s)$,
where the dash denotes differentiation w.r.t. $s$. The extended
action $S_e$ which governs the dynamics of $q(s),t(s)$ is then given
by
\begin{eqnarray}\label{Gl:OneDimensionalExtendedAction}
S_e\;:=\;\int ds\;L_e\big(q(s),t(s),q'(s),t'(s)\big).
\end{eqnarray}

\noindent The Euler-Lagrange equation for $t(s)$ defined by
(\ref{Gl:OneDimensionalExtendedAction}) can easily be shown to be automatically satisfied, due to the chain rule and using the equation for $q(s)$ \cite{bahrdittrich2}. As a result, the system is underdetermined, and the boundary conditions are not sufficient to determine $t(s), q(s)$ uniquely. Rather there is a (gauge) freedom which is exactly in the free choice of reparametrization $t(s)$. With a choice of $t(s)$, the equation for $q(s)$ can be uniquely solved, and reproduces the dynamics of
the action (\ref{Gl:OneDimensionalAction}). 
%

To discretize the system, consider the functions $q_n:=q(s_n)$ and
$t_n:=t(s_n)$ at only finitely many values
$s_{initial}=s_0<s_1<\ldots<s_N=s_{final}$ for $s$. The $t_n, q_n$ are the
configuration variables of the discretized system, and their
dynamics is governed by a discretization of
(\ref{Gl:OneDimensionalExtendedAction}), the \emph{discrete extended
action}
\begin{eqnarray}\label{Gl:OneDimensionalExtendedAction_Discrete}
S_{de}\;:=\;\sum_{n=0}^{N-1}\;L_n\;\,\,(t_{n+1}-t_n)\;:=\;\sum_{n=0}^{N-1}L\left(q_n,\;\frac{q_{n+1}-q_n}{t_{n+1}-t_n}\right)(t_{n+1}-t_n) \, .
\end{eqnarray}

 Whereas in the continuum the equation following from the variation of $t(s)$ is automatically satisfied if the equation following from the variation of $q(s)$ is, the equivalent statement does not hold in general for the discrete theory. Rather the equations from varying $t_n$ constitute a discretization of the chain rule, which is however not automatically satisfied as derivatives are approximated by finite difference quotients \cite{bahrdittrich2}. As a consequence these equations have to be solved for the $t_n$, leading to unique solutions also for the $q_n$.
 Hence the reparametrization invariance of the continuum theory -- which would correspond to a free choice of the $t_n$ -- is broken by the discretization.


In classical theories, gauge symmetries are usually accompanied by singular Hessians, i.e. the fact that the matrix of second derivatives of the action, evaluated on a solution, has zero eigenvalues. The eigenvectors to these zero eigenvalues correspond to infinitesimal gauge transformations. In contrast the Hessian of the discrete extended action (\ref{Gl:OneDimensionalExtendedAction_Discrete}) has only non--vanishing eigenvalues in generic cases, i.e. for a particle moving in a potential \cite{bahrdittrich1}.

Nevertheless in the continuum limit the eigenvalues of the Hessian, corresponding to the gauge degrees of freedom, tend to zero, while the others, which correspond to the physical degrees of freedom of the continuum theory\footnote{i.e. the functional dependence $q(t)$ which does not change under reparametrizations}, stay finite. 
 The corresponding modes
 (i.e. the corresponding variables in the linearized theory, which is determined by the Hessian)
 should therefore exhibit very different dynamical behaviour. Indeed consider the contributions to the action from only  the kinematical term
\begin{eqnarray}
S^{kin}_{de}:=\sum_{n=0}^{N-1} \frac{m}{2} \frac{(q_{n+1}-q_n)^2}{(t_{n+1}-t_n)^2}(t_{n+1}-t_n) \quad .
\end{eqnarray}
The Hessian, i.e. the matrix of second derivatives of $S^{kin}$ with respect to the variables \\$(t_1,q_1,\ldots,t_{N-1},q_{N-1})$, has $(N-2)$ null vectors $g^m,\,m=1,\ldots,N-2$ even off--shell, i.e. before invoking the equations of motions.  The components $g^m_k=(g^m_{t_k},g^m_{q_k})$ are given by
\begin{eqnarray}\label{pseudogauge}
g^m_k = ( \delta^m_k  - \frac{v_m-v_{m-1}}{v_{m+1}-v_m} \delta^m_{k-1}\,, \, \,\,   v_{m-1}\delta^m_k -v_{m+1}\frac{v_m-v_{m-1}}{v_{m+1}-v_m} \delta^m_{k-1}) \quad
\end{eqnarray}
where $v_m:=\frac{q_{m+1}-q_m}{t_{m+1}-t_m}$ denotes the discretized velocities. For the free particle we have -- using the equations of motions -- $v_m=v=$const.  and we actually obtain $(N-1)$ exact gauge modes $g^m_k=(\delta^m_k,v\delta^m_k)$. For the general case with non--constant potential the pseudo gauge modes (\ref{pseudogauge}) have a vanishing kinematical term and get a non--vanishing contribution only via the potential. Hence their dynamical properties, for instance correlation functions, differ from those of the remaining modes with a non--vanishing kinematical term. In Regge gravity a linearization of the theory around a curved background should also lead to pseudo gauge modes \cite{bahrdittrich1} and it would be valuable to investigate whether these actually  propagate or not.

For an interpretation of the discrete theory, one therefore needs an analysis which is able to distinguish between those degrees of freedom which survive the continuum limit, and those which are approximate gauge degrees of freedom.  Otherwise one might draw conclusions, that do not correspond to any features of the continuum theory. The so--called problem of observables \cite{carlo, bd}, that is the problem of identifying the physical degrees of freedom, still exists.\footnote{In a sense it is even more severe as there is not an exact notion of gauge in the discrete theory and therefore no exact notion of observables yet.}

\subsubsection{Regaining reparametrization invariance}

Although the gauge symmetries are broken in the discretized theory these have to be restored in the continuum limit in order to recover the original theory. One method to establish the symmetries also in the discrete theory is to discretize the theory in a way, such that it reflects exactly the dynamics and hence gauge symmetries of the continuum theory. Wilson's renormalization approach \cite{wilson, perfect} provides us with the necessary technique: if we start with a very fine grained lattice and integrate out\footnote{In the classical theory this just means to solve the equations of motions for the fine grained degrees of freedom.} the fine grained degrees of freedom the resulting effective theory -- although defined on the coarse grained lattice -- delivers the predictions of the fine grained theory.  Taking the limit of the fine graining to the continuum we obtain a theory on a lattice that exactly reflects the predictions of the continuum theory. The corresponding action is called perfect. For the one--dimensional systems in the previous section this `blocking from the continuum' \cite{wisefischer} leads to a sum of Hamilton--Jacobi functionals
\begin{eqnarray}\label{exact}
S_{exact}\;=\;\sum_{n=0}^N S_{HJ}^{(n)}(t_n,q_n,t_{n+1},q_{n+1})\;=\;\sum_{n=0}^N\int_{s_n}^{s_{n+1}}ds\;L_e(t(s),q(s)).
\end{eqnarray}
where in the last expression the action is evaluated on the solution of the (continuum) equations of motions with boundary conditions $(t_n,q_n,t_{n+1},q_{n+1})$.  Solutions of the discrete theory defined by the exact action (\ref{exact}) can be obtained by taking any of the continuum solutions $(t(s),q(s))$ with the same boundary conditions and extracting the discrete values $(t(s_n),q(s_n))$. As there are many continuum solutions to the same boundary conditions we also obtain many discrete solutions. Hence this discrete theory exhibits gauge symmetries.




\subsection{$3D$ Regge Gravity with a cosmological constant}

An example, in which the breaking of the symmetry upon discretization can be observed as well, is $3D$ Regge calculus with a cosmological constant. 
%
 For a triangulation $\mathcal{T}$ of a $3D$-manifold with edges $E$, triangles $T$ and tetrahedra $\Sigma$ the Regge action
\begin{eqnarray}\label{Gl:ReggeAction3DWithLambda}
S_{\mathcal{T}}\;=\;\sum_{E}L_E\epsilon_E\;-\;\Lambda\sum_{\Sigma}V_\Sigma
\end{eqnarray}

\noindent has a Hessian with nonzero determinant if and only if $\Lambda=0$.\footnote{ For zero cosmological constant the gauge symmetry of the continuum is not broken -- the reason is that the solutions of the discrete theory (locally flat geometries) coincide with thus of the continuum.} Here $L_E$ denotes the length of $E$, $V_\Sigma$ the volume of $\Sigma$ and $\epsilon_E$ is the deficite angle at $E$.
As soon as $\Lambda\neq 0$, the Hessian has no longer a zero determinant \cite{dittrich08}, i.e. the term with the cosmological constant explicitly breaks the gauge symmetry.
It is possible to construct a perfect action that is a function only of the edge lengths $L_E$ of a given triangulation $\mathcal{T}$, but nevertheless gives the exact solutions of the continuum theory. This action exhibits gauge symmetries, that correspond to displacing the vertices of the triangulation and can be seen as a discrete analogue of the diffeomorphism symmetry.

To construct an improved action, the triangulation $\mathcal{T}$ is refined to a triangulation $\tau$ with edges $e$ and tetrahedra $\sigma$. For fixed $L_E$ the Regge equations for the lengths $l_e$ are solved, under the conditions that $\sum_{e\subset E}l_e=L_E$. These condition can be added o the Regge action via Lagrange multipliers $\alpha_E$, which results in
\begin{eqnarray}\label{Gl:ReggeAction3DWithConstraints}
S_\tau\;=\;\sum_el_e\epsilon_e\;-\;\Lambda\sum_{\sigma}V_\sigma\;+\;\sum_E\alpha_E\left(L_E-\sum_{e\subset E}l_e\right).
\end{eqnarray}
 The value of the Regge action $S_\tau$ on a solution
$
S_{\mathcal{T},\tau}\;=\;{S_{\tau}}_{\|_{\frac{\partial S_{\tau}}{\partial l_e}=\frac{\partial S_\tau}{\partial \alpha_E}=0}}
$
%
 is a function of the $L_E$ only, hence it ``lives" on the triangulation $\mathcal{T}$, but reflects the dynamics of the fine-grained triangulation $\tau$. It can be shown to be
\begin{eqnarray}
S_{\mathcal{T},\tau}\;=\;\sum_EL_E\alpha_E\;+\;2\Lambda\sum_\Sigma V_\Sigma
\end{eqnarray}

\noindent where $V_\Sigma:=\sum_{\sigma\subset\Sigma}V_\Sigma$, and the Lagrange multipliers $\alpha_E$ have to be determined by the equation of motion defined by (\ref{Gl:ReggeAction3DWithConstraints}). 
In the continuum limit one can show that
 the $\alpha_E$ and $V_\Sigma$ converge to deficit angles $\epsilon_E^{(\kappa)}$ and volumes $V_\Sigma^{(\kappa)}$ in tetrahedra of constant curvature $\kappa=\Lambda$, having the edge lengths $L_E$. 
 The \emph{perfect action}, which is the limit of $S_{\mathcal{T},\tau}$ for infinitely fine $\tau$, is therefore given by
\begin{eqnarray}\label{Gl:PerfectAction3D}
S_{\mathcal{T},*}\;:=\;\lim_{\tau\to\infty}S_{\mathcal{T},\tau}\;=\;\sum_EL_E\epsilon_E^{(\Lambda)}\;+\;2\Lambda\sum_\Sigma V_\Sigma^{(\Lambda)},
\end{eqnarray}

\noindent i.e. the Regge action with cosmological constant, for tetrahedra of constant curvature $\kappa=\Lambda$. Since the equations of motion $\partial S_{\mathcal{T},*}/\partial L_E=\epsilon_E^{(\Lambda)}=0$ demand the vanishing of the deficit angles, the solutions describe $3D$ geometries with constant curvature $\Lambda$. These solutions coincide with those of $3D$ continuum GR with a cosmological constant. The $L_E$ are not uniquely determined by the equations of motion, but\footnote{similar to the case of $\Lambda=0$} there are three gauge degrees of freedom per vertex. These correspond to the vertex displacement symmetry, which is a result of the non-uniqueness of triangulation of a constantly curved manifold with constantly curved tetrahedra. Hence, the perfect action (\ref{Gl:PerfectAction3D}), although being discrete in the sense that it is defined on the triangulation $\mathcal{T}$,  mirrors the correct gauge symmetry of the continuum theory.

\subsection{Higher dimensions}

In $4D$ Regge Gravity (with or without cosmological constant) the gauge symmetry of the continuum theory is also broken. With the exception of $\Lambda=0$ and the special case of a flat solution\footnote{Which again is exactly the case in which the discrete and the continuum dynamics produce the same solution, i.e. locally flat metrics.}, 
 the length assignment in general is unique for given boundary data. For generic solutions, the Hessian of the theory has eigenvalues which tend to zero at most with $\epsilon^2$, where $\epsilon$ is the order of magnitude of the deficit angles at the triangles \cite{bahrdittrich1, ruthrocek}.

%
%
%
%
%
Since the Einstein equations are much more complicated in more than three dimensions, the analysis of the improved an perfect actions are much more involved. A surprising amount of observations can be obtained right away, however.

In particular, for $4D$ the improved actions can be defined in a similar manner as for $3D$\footnote{The analysis for $4D$ can be easily extended to any higher dimension. Also, for higher dimensions we include the cosmological constant term, for the sake of generality.} by subdividing a triangulation $\mathcal{T}$ with edges $E$, triangles $T$ and simplices $\Sigma$ into a finer triangulation $\tau$ with edges $e$, triangles $t$ and simplices $\sigma$, and solving the Regge equations for $\tau$. In $4D$, however, it proves to be more convenient to keep the area $A_T$ of the triangles $T$ fixed, rather than the edge lengths $L_E$. The reason is that this keeps the form of the action invariant.  The Regge action to be varied is then
\begin{eqnarray}\label{Gl:ReggeAction4DWithConstraints}
S_{\tau}\;=\;\sum_ta_t\epsilon_t\;-\;\Lambda\sum_\sigma V_\sigma\;+\;\sum_T\alpha_T\left(A_T\,-\,\sum_{t\subset T}a_t\right)
\end{eqnarray}
\noindent where the Lagrange multipliers $\alpha_T$ are introduced to ensure $\sum_{t\subset T}a_t=A_T$. The improved action can then be computed to be
\begin{eqnarray}\label{Gl:ImprovedAction4D}
S_{\mathcal{T},\tau}\;=\;{S_{\tau}}_{\|_{\frac{\partial S_{\tau}}{\partial l_e}=\frac{\partial S_\tau}{\partial \alpha_H}=0}}
\;=\;\sum_TA_T\alpha_T\;+\;\Lambda\sum_\Sigma V_\Sigma
\end{eqnarray}
\noindent just as in $3D$. The improved action $S_{\mathcal{T},\tau}$ is a function of the areas $A_T$, as are the $\alpha_T$ and $V_\Sigma$, which have to be obtained by solving the Regge equations. As in 3D there is a formal agreement with the Regge action for curved simplices in $4D$ with constant curvature $\Lambda=3\kappa$. 


There are some difficulties however. One is that the areas usually over--determine a triangulation, that is areas coming from a consistent length assignment have to satisfy constraints \cite{areaangle}. These constraints depend on whether one has flat or curved simplices \cite{bahrdittrich3}, so it is not even obvious which constraints to apply. Hence it is not clear for which sets of $\{A_T\}_{T\in \mathcal{T}}$ a solution to the equations of motion defined by (\ref{Gl:ReggeAction4DWithConstraints}) exist and moreover the interpretation of the $\alpha_T$ as deficit angles is not necessarily valid.

However, assuming that the  perfect limit exists, one can derive several properties of the perfect action, e.g. the equation of motion $\frac{\partial S_{\mathcal{T},*}}{\partial A_T}=\alpha_T$, as well as a first order ODE w.r.t. $\Lambda$
\begin{eqnarray}\label{Gl:EOMForImprovedActionCurvedSimplices}
S_{\mathcal{T},*}\;+\;\frac{1}{3}\Lambda\frac{\partial
S_{\mathcal{T},*}}{\partial \Lambda}\;=\;\sum_{T}\alpha_T A_T.
\end{eqnarray}

\noindent  This equation can be used to show that e.g. the improved actions defined with flat and with curved simplices converge to the same perfect action in the continuum limit $\tau\to\infty$, and to compute the perfect action explicitly for certain values of the $A_T$. In those special cases one can prove that the perfect action regains the correct gauge symmetry of the continuum theory.

Another possibility is to compute the perfect action perturbatively in an expansion around a flat background as is discussed in the last section.

\section{Repercussions for the canonical theory}

Gauge symmetries of the action lead to constraints in the canonical formulation of the theory, i.e. restrictions on the configurations and momenta at a given time. General Relativity is even a totally constraint system, meaning that the full dynamics is determined by the constraints.

We have seen that discretizations generically seem to break diffeomorphism invariance, hence the constraints will be affected too. Attempts to discretize directly the constraints lead to algebras that are not closed (that is second class constraints), hampering a consistent definition of the dynamics. Alternatively one can derive the canonical theory from the discrete action \cite{bahrdittrich1}. In this case the canonical formulation exactly mirrors the symmetry content of the covariant theory: If there are exact symmetries we obtain exact (first class) constraints. Hence constructing an action with exact symmetries will also solve the old problem of obtaining a discrete canonical theory with a consistent constraint algebra! A consistent constraint algebra can  be obtained for cases in which the symmetries are already exact, that is for 3D gravity \cite{waelbroeck} and for the `flat sector' of 4D gravity \cite{jimmy}, i.e. for spatial triangulations, such as the surface of a 4--simplex, that only allow for flat 4D solutions.

 If there are only approximate symmetries in the action, one obtains so--called pseudo constraints \cite{gambini,bahrdittrich1}, i.e. constraints that do not only involve the canonical data of one time step but also data of the next time step. Hence these are rather proper equations of motion. However, the dependence on the data of the next time step is very weak if the symmetries are only weakly broken. Indeed if we expand for instance Regge gravity on a flat background, then the linearized theory still exhibits exact (linearized) gauge invariance \cite{ruthrocek,bdlfss}. Correspondingly one can find linearized Abelian constraints in the canonical formulation \cite{toappear}. Only  at higher order do these constraints turn into pseudo constraints, i.e. acquire terms that depend on the data at the next time step.

\section{Conclusion and Outlook}


In summary although discrete theories of 4D gravity seem to generically break diffeomorphism invariance there is a possibility to construct discretizations that display exact diffeomorphism invariance. Requiring diffeomorphism invariance might avoid discretization and quantization ambiguities \cite{lost,ashtekarpaper, kuchar}.

Even if it is calculationally not possible to obtain the exact perfect action, a systematic investigation of the approximate symmetries and constraints under refinement of the triangulation\footnote{Which can be seen as renormalization group flow.}  is paramount for the question of the fate of diffeomorphism symmetry in the discretized setting.
This will help to derive quantization conditions that allow to regain the symmetries in the continuum limit after quantization.  To this end one might need to investigate the fate of the pseudo gauge modes upon quantization .

Moreover it would be valuable to have structural information about perfect actions for interacting diffeomorphism invariant field theories, for instance about their locality properties. Here a perturbative approach might be useful, i.e. starting from an expansion of the Regge action on a flat background. Whereas the linearized theory exhibits gauge symmetries these are broken by the higher order interaction terms. Here the question is whether we can regain these symmetries order by order \cite{wip}. As the perfect action reflects the dynamics of the continuum theory this would also allow us to make connections between continuum approaches to quantum gravity and inherently discrete ones.



\end{document}